\documentclass[epsf,twocolumn,showpacs,preprintnumbers]{revtex4}
\usepackage{graphics}
\usepackage{graphicx}% Include figure files
\usepackage{dcolumn} % Align table columns on decimal point
\usepackage{bm}
\usepackage{epsfig}
\begin{document}
\title{Study of Correlation Effects in the \\
                High Formal Oxidation
       State Compound  Sr$_2$CoO$_4$ } 
\author{K.-W. Lee and W. E. Pickett} 
\affiliation{Department of Physics, University of California, Davis, 
 California 95616, USA}
\date{\today}
\begin{abstract}
Two recent reports confirm that the newly synthesized Sr$_2$CoO$_4$ 
(formal oxidation state Co$^{4+}$)
shows a high Curie temperature ($\sim 250$ K), but they report different moments 
of 1.8 $\mu_B$ and 1 $\mu_B$ per Co.  Using both commonly used functionals in the
correlated band approach (LDA+U) as well as the local density approximation (LDA),
the combined effects of correlation and hybridization with O $2p$ states are
calculated and analyzed.  
Sr$_2$CoO$_2$ is already ferromagnetic
within LDA ($M$=1.95 $\mu_B$).  Increasing $U$ from zero, the two LDA+U schemes
affect the moment oppositely out to a critical value
$U_c$=2.5 eV, at which point they transform
discontinuously from different states to the {\it same} 
large $U$ state.  Fixing $U$ at $U_c$, fixed
spin moment calculations show similar behavior out to a 
minimum at 1$\mu_B$ (a half metallic state), beyond which the
fully-localized-limit scheme jumps to a state with energy minimum very near 2$\mu_B$
(very close to the LDA moment).   Although the energy minima occur very near
integer values of the moment/Co ($1\mu_B, 2\mu_B$), the strong $3d-2p$ mixing 
and resulting $3d$ orbital occupations seem to
preclude any meaningful $S=\frac{1}{2}$ or $S=1$ assignment to the Co ion.
\end{abstract}
\maketitle

\section{introduction}
%LSCoO4
First row ($3d$) transition metal ions in high oxidation states  
have been of interest for some time, due to their competing and delicate
spin states.  There are still several cases where
behavior is not understood, indeed sometimes (due to sample questions
arising from difficulty in synthesis) the data is not unambiguous.  The best
known, and most thoroughly studied, example is that of 
the quasi-two-dimensional (2D) cuprates,
which when hole-doped become high temperature superconductors.  In their
undoped state they are Cu$^{2+}$-based antiferromagnetic insulators.
Hole doping drives the Cu oxidation state toward the unstable (practically
non-existent) Cu$^{3+}$ state.  At a doping level of 0.15-0.20 holes/Cu
optimal superconductivity is reached.  They can be doped beyond that level,
when they become conventional metallic Fermi liquids.  Upon hole doping
the `oxidation state' designation must be interpreted with care, since
it is clear that the holes go onto the oxygen ions to a large degree.

Another high-oxidation-state ion is Co$^{4+}$.  This ion has recently
gained wide exposure due to the unusual properties of Na$_x$CoO$_2$,
including the discovery\cite{takada} that this system becomes superconducting (4.5 K)
around $x$=0.3 when it is hydrated.  This value of $x$ suggests that
the Co ion is 70\% of the way to being Co$^{4+}$, or alternatively,
70\% of the ions are Co$^{4+}$ while 30\% are Co$^{3+}$.  This system is
however metallic for all $x$ except for Na ion/orbital/spin/charge
ordering\cite{foo} precisely at $x$=0.5 (precisely what is responsible is not yet clear).
There is magnetism and correlated electron behavior for $x >
0.5$ (the Co$^{3+}$ end) while for $x < 0.5$ the materials appear to be
weakly correlated nonmagnetic metals.  While there has been much expectation
that the (metastable) endpoint member CoO$_2$ (nominally Co$^{4+}, d^5$) is
a Mott insulator, the evidence is 
that it remains a nonmagnetic metal.\cite{tarascon}  
Calculations indicate\cite{DFTcalcs} that, as in the 
cuprates, upon hole doping from the Co$^{3+}$ end, much of the charge
difference occurs on the O ion.

The La$_{2-x}$Sr$_x$CoO$_4$ system (LSCoO) with two-dimensional layered
K$_2$NiF$_4$ structure has become interesting 
because of its magnetic and electrical properties.   In this system 
the formal oxidation state Co$^{(2+x)+}$ near the La end is not unusually high,
so its behavior might be expected to be readily understandable.
With increasing Sr concentration $x$, LSCoO shows a structural transition 
from orthorhombic to tetragonal (i.e., a lattice constant $b$$\rightarrow$$a$)
around $x$=0.5,  
with enhanced two-dimensional electronic properties.\cite{matsuura}
The structural transition may be connected  to an antiferromagnetic-ferromagnetic 
transition (the maximum Curie temperature $T_C$=220 K 
at $x$=0.9),\cite{furukawa} accompanying a magnetic change that has been
interpreted in terms of a Co$^{3+}$ spin-state
transition from high spin to intermediate spin configurations
around $x$=0.7.\cite{moritomo}
There is another suggestion that at $x$=1 the system is a high-spin, low-spin 
charge-ordered state.\cite{jwang}
However, there is no agreement amongst the measurements on
the magnetic behavior \cite{furukawa,sanchez,demazeau,liu} 
nor metallic behavior\cite{furukawa,rao}.
The end member $x$=0 is an antiferromagnetic (AFM) insulator 
with Neel temperature $T_N$=275 K.\cite{yamada} 

%Sr2CoO4
Although earlier studies were confined to the range  below $x$=1.4,  
recently the end member Sr$_2$CoO$_4$, formally Co$^{4+}$, was synthesized by
Matsuno {\it et al.}\cite{matsuno,matsuno2} and by 
Wang {\it et al.}\cite{xwang,xwang2},
and characterized as FM with 
high Curie temperature $T_C$$\approx$250 K. 
However, their differing synthetic methods have led to different
properties.
Matsuno {\it et al.} synthesized a single-crystalline thin film
using pulsed-laser deposition, 
while Wang {\it et al.} produced polycrystalline samples under 
high pressure, high temperature conditions.
The former shows metallic $T$-dependent resistivity below $T_C$, 
although it has definitely higher resistivity of 
order of 10$^{-4}-10^{-3}$ $\Omega$ cm at low $T$ than 
in a typical metal.
The pressure-synthesized samples show nearly temperature independent resistivity
(perhaps due to polycrystallinity), but
with similar magnitude.
The more confusing difference is in the observed saturation magnetic
moment, 1.8 $\mu_B$ and 1 $\mu_B$/Co respectively.
Nevertheless, the sample with the smaller ordered moment has been 
observed\cite{xwang} effective (Curie-Weiss) moment $p_{eff}$= 
3.72 $\mu_B$, characteristic of a
much higher $S=3/2$ spin configuration.  Such a moment would suggest a
much higher ordered moment, $<S_x> \sim 3\mu_B$.

%LSCoO3
The related perovskite system La$_{1-x}$Sr$_x$CoO$_3$ (formally Co$^{(3+x)+}$)
has been studied
for some time.\cite{tekeda,mathi,potze,korotin,takahashi,nekrasov,maris}
Its magnetic properties are also altered by Sr concentration $x$:
nonmagnetic for $x<$ 0.05, spin glass for 0.05 $\le x <$ 0.2,
and ferromagnetic above $x$=0.2.
For the end member, SrCoO$_3$ (``Co$^{4+}$'') shows metallic conductivity and has
magnetic moments 1.25 and 0.1 $\mu_B$ for Co and O respectively,\cite{tekeda}
whereas the ground state of LaCoO$_3$ is a nonmagnetic band insulator.
Its excitations have been interpreted in terms of a locally orbitally-ordered
excited state.\cite{maris,phelan}

In this paper, we look in some detail at the electronic and magnetic structure 
of Sr$_2$CoO$_4$ both as uncorrelated, using the local
density approximation (LDA), and viewing effects of correlation as
described by the LDA+Hubbard $U$ (LDA+U) 
method.  Both commonly used LDA+U functionals are employed, and their
results are compared and contrasted.
A number of unusually rich magnetic phenomena arise (including a half metallic
phase), reflecting the strong hybridization with O $2p$ states that complicates
the accommodation of correlation effects on the Co ion.

\section{Structure and calculation}
\begin{figure}[tbp]
{\resizebox{5cm}{10cm}{\includegraphics{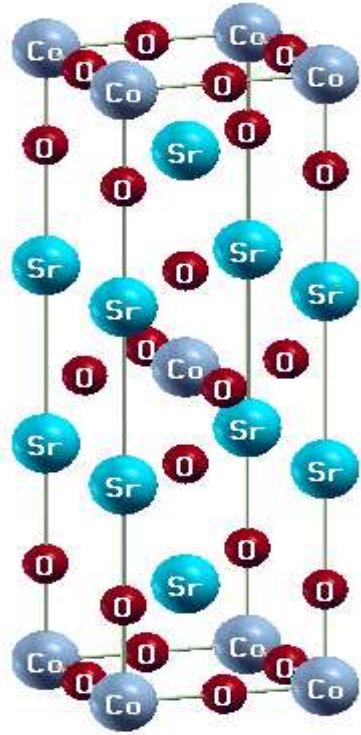}}}
\caption{(Color online) Crystal structure of Sr$_2$CoO$_4$, 
 a body-centered-tetragonal type. The planar O-Co bond length is 
 1.878~\AA, about 6\% shorter than the apical O-Co bond length.
 Lattice constants used here are $a$=3.755~\AA~ and $c$=12.6~\AA.}
\label{cell}
\end{figure}

Sr$_2$CoO$_4$ has the bct 
K$_2$NiF$_4$-type structure, space group $I4/mmm$ (No. 139),
pictured in Fig. \ref{cell}. The 
Co - planar O (PO) bond length (1.878~\AA) is shorter by about 6\% than 
that of Co - apical O (AO) (1.988~\AA).
The distorted CoO$_6$ octahedron, elongated along $c$-axis, leads to
crystal field splitting of $t_{2g} \rightarrow$ E$_g$($d_{xz}, d_{yz}$)
+ B$_{2g}$($d_{xy}$) and
$e_g\rightarrow$ A$_g$($d_{3z^2-r^2}$) + B$_{1g}$($d_{x^2-y^2}$) states. 
We have used the lattice constants $a$=3.755~\AA, $c$=12.6~\AA, and
apical O (0.1578) and Sr (0.3544) internal parameters optimized by 
Matsuno {\it et al.},\cite{matsuno} which are consistent 
with the experimentally measured values by Wang {\it et al}.\cite{xwang}

Our calculations were carried out within the local density approximation (LDA)
and LDA+U approaches with the full-potential nonorthogonal local-orbital 
(FPLO) method.\cite{fplo1} 
Both commonly used schemes of the LDA+U method were employed so comparisons
of the predictions could be made.
Both forms have the same Hubbard-like density-density interaction
\begin{eqnarray}
 &E_{U}&= \frac{1}{2}\sum_{m\sigma \neq m'\sigma '}
  [U_{mm'}-\delta_{\sigma,\sigma '}J_{mm'}] n_{m\sigma} n_{m'\sigma '},
\end{eqnarray}
where $\{n_{m\sigma}\}$ is a site occupation set and
$J$ is intra exchange integral.  The form displayed here is schematic in
the sense that it does not display all of the indices involved in the
full coordinate-system-independent form that is implemented in the code.
The two LDA+U approaches differ
only in the method of treatment of
the ``double counting" term, intended to subtract out the shell-averaged
interaction that has already been included in LDA.
One choice is the so-called ``around mean field" (AMF) scheme, which is 
expected to be more
suitable when the on-site Coulomb repulsion $U$ is not so strong.\cite{ldau1}
The other choice, called the ``fully localized limit" (FLL)
(also called ``atomic limit"), is more appropriate for large $U$ systems.\cite{ldau2}
The methods to treat the double counting problem in the both schemes 
are given by
\begin{eqnarray}
 E^{dc}_{AMF}&=& \frac{1}{2}\sum_{m\sigma \neq m'\sigma '}
  [U_{mm'}-\delta_{\sigma,\sigma '}J_{mm'}]~\bar{n}~\bar{n},\\ \nonumber
 E^{dc}_{FLL}&=& \frac{1}{2}\sum_{m\sigma \neq m'\sigma'}
  [U_{mm'}-\delta_{\sigma,\sigma '}J_{mm'}] 
     \bar{n}_{\sigma} \bar{n}_{\sigma '},
% E_{U}^{AMF}&= \frac{1}{2}\sum_{m\sigma}\{U(N-n_\sigma)
%         -J(N_\sigma-n_\sigma)\}N_\sigma   \\ \nonumber
% E_{U}^{FLL}&= \frac{1}{2}\sum_{m}\{UN(N-1)
%         -J\sum_{\sigma}N_\sigma(N_\sigma-1)\}          \\
%&N_\sigma&= \sum_{m} n_{m\sigma}~,~~ N = \sum_{\sigma}N_\sigma
\end{eqnarray}
where $\bar{n}$ is the shell-averaged occupation, while $\bar{n}_{\sigma}$
is its spin-decomposed analog.  These double-counting terms can be 
written in other forms that emphasize other aspects of the 
interaction.\cite{dudarev}  However, this form is illustrative because it 
emphasizes that the difference lies in the magnitude of the (self-consistent)
atomic moment.
Clearly it is only the spherically averaged values of $U$ and $J$
that enter the double-counting terms.  The difference between the two forms
is that the double-counting term, and the resulting potential, 
includes a spin dependence in the
FLL form.

The fully relativistic scheme of FPLO\cite{fplo2} was used  when
spin-orbit coupling (SOC) in LDA was necessary.
The choice of basis orbitals were Sr $(4s4p)5s5p4d$, Co $(3s3p)4s4p3d$, 
and O $2s2p(3s3p3d)$. 
(The orbitals in parentheses indicate semi-core or polarization orbitals.)
The Brillouin zone was sampled by a regular mesh 
containing up to 641 irreducible {\bf k} points for LDA and LDA+U, 
and 1639 irreducible {\bf k} points 
for SOC and fixed spin moment (FSM)\cite{fsm} calculations
that require more carefully treatment near the Fermi level ($E_F$).

\section{Uncorrelated Treatment}
\subsection{Electronic structure}
\label{ldaA}

\begin{figure}[tbp]
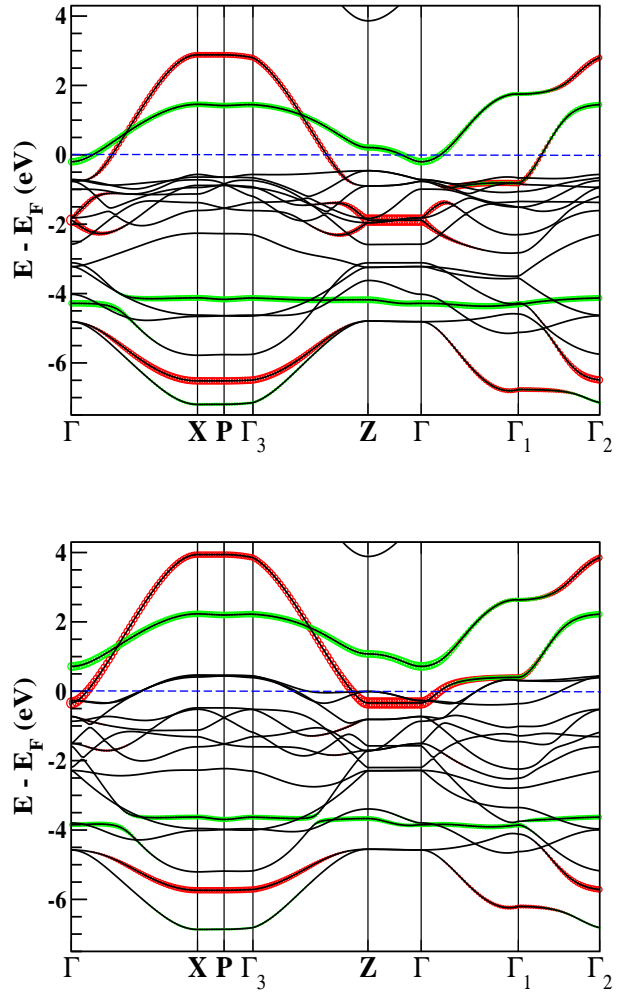

{\resizebox{8cm}{6cm}{\includegraphics{Fig2a.eps}}}
\vskip 11mm
{\resizebox{8cm}{6cm}{\includegraphics{Fig2b.eps}}}
\caption{(Color online) FM LDA majority (upper) and 
 minority (lower) band structures along symmetry directions.
 The thickened (and colored) lines emphasize Co $d_{3z^2-r^2}$ 
 (green or light) and $d_{x^2-y^2}$ (red or black) characters,
 which form bonding and antibonding bands with apical and 
 in-plane O $p_\sigma$
 states respectively.
 The symmetry points for the body-centered-tetragonal structure
 follow the Bradley and Cracknell notation as given in Fig. \ref{FS}. 
 The dashed horizontal line denotes the Fermi energy.}
\label{fmband}
\end{figure}

\begin{figure}[tbp]
\rotatebox{-90}{\resizebox{7cm}{8cm}{\includegraphics{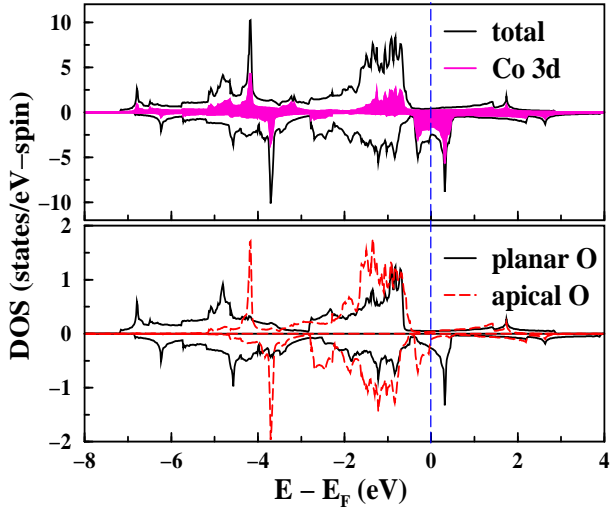}}}
\caption{(Color online) Total and atom-projected densities of states 
 for LDA FM calculation.
 Near $E_F$, there is a van Hove singularity in the
 minority channel.
 While AO is almost fully occupied, the minority of PO is partially
 occupied, resulting in large magnetic moment for PO (for details,
 see text).
 Additionally, the band width of AO is by 40\% less than that of PO.
 The DOS at $E_F$ $N(0)$ is 2.90 states/eV per both spins.}
\label{dos}
\end{figure}

\begin{figure}[tbp]
\rotatebox{-90}{\resizebox{7cm}{8cm}{\includegraphics{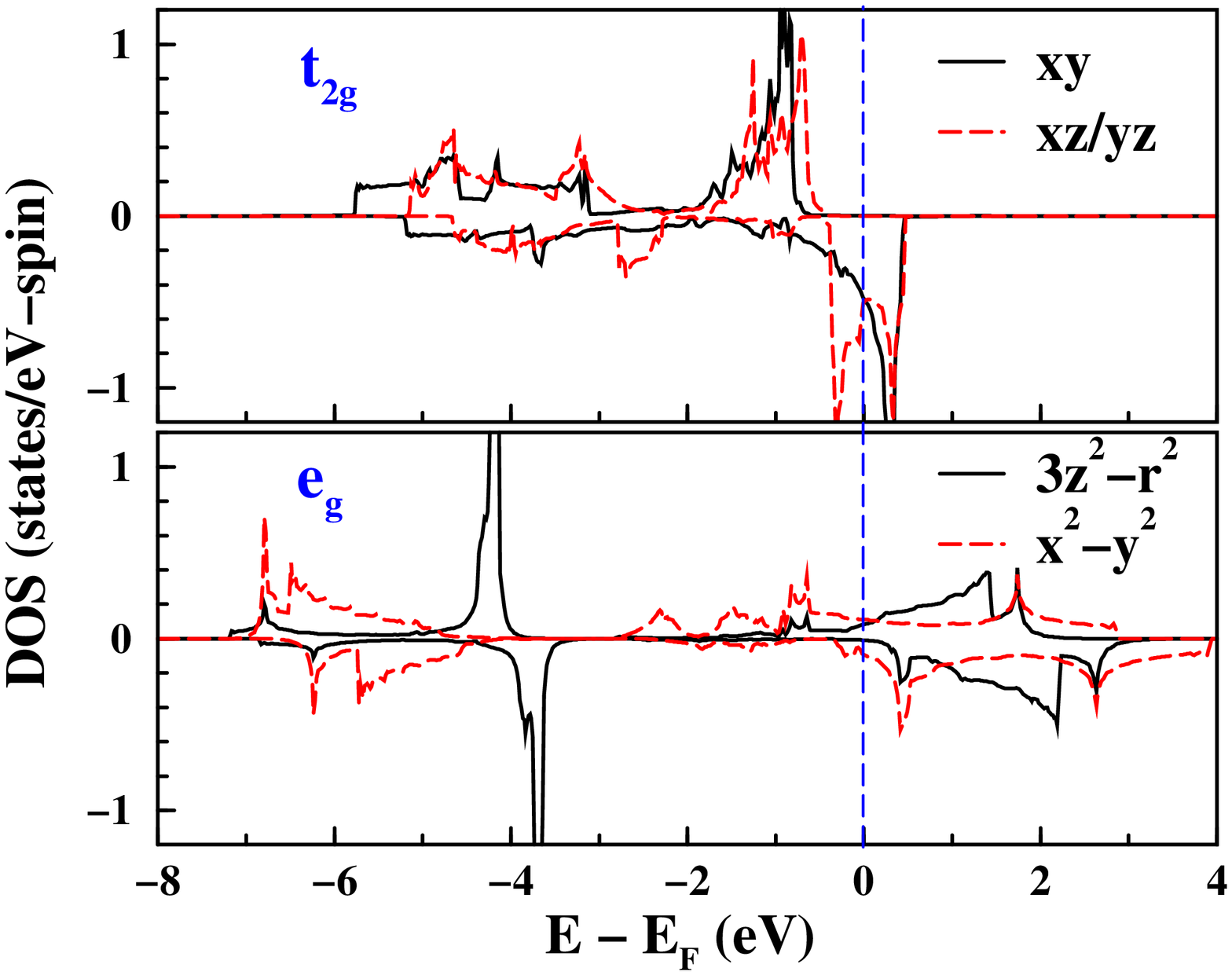}}}
\rotatebox{-90}{\resizebox{7cm}{8cm}{\includegraphics{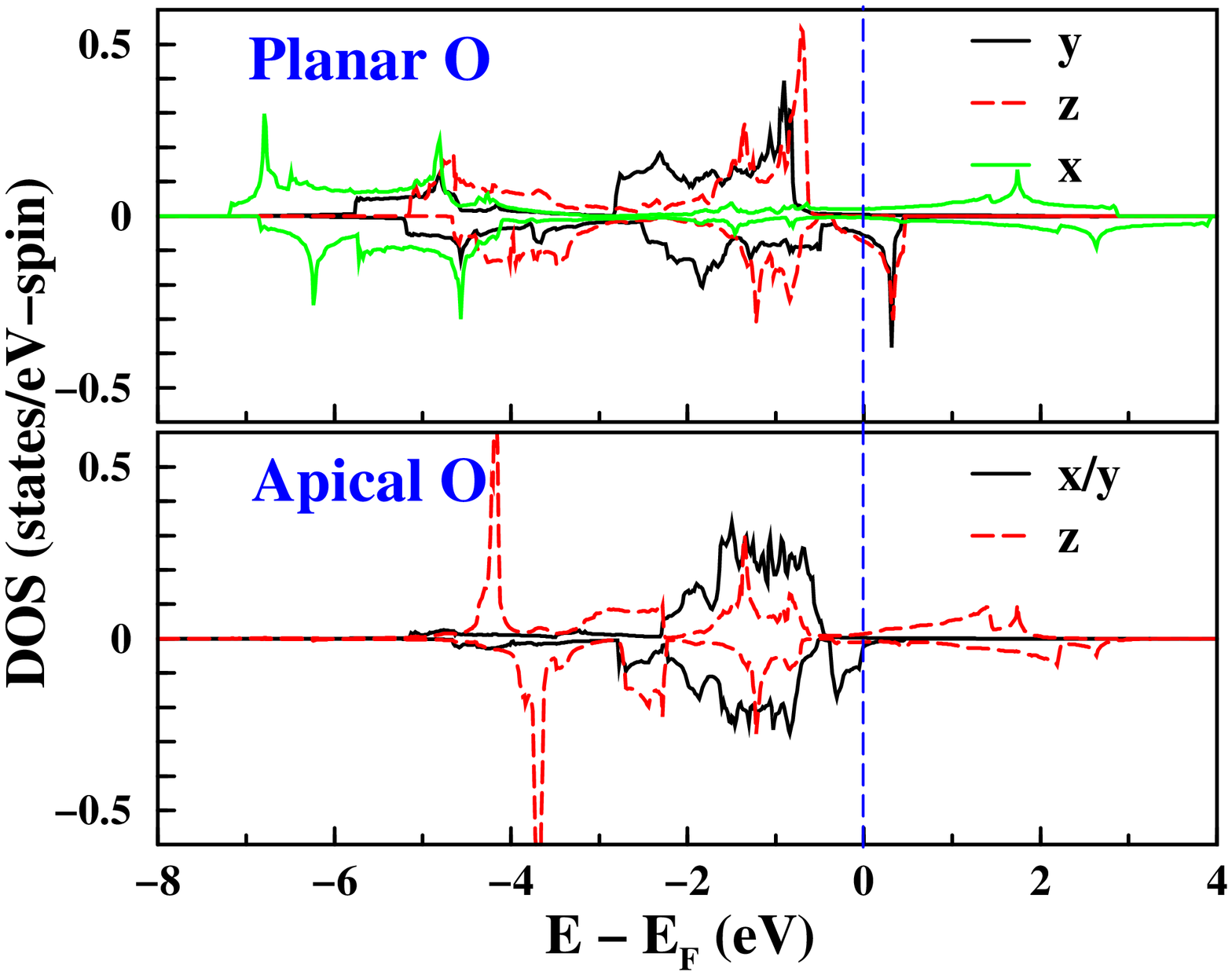}}}
\caption{(Color online) Orbital-projected densities of states for
 Co $3d$ (upper) and O $p$ states (lower) in LDA FM
 calculation. The $d_{xz}$ and $d_{yz}$ states are degenerate. 
  The crystal field splitting of 2 eV between $t_{2g}$ and $e_g$ manifolds
 is only a little higher than the exchange splitting 1.3 eV.
 The PO $p_x$ and AO $p_z$ are the $\sigma$-orbital.}
\label{pdos}
\end{figure}

The FM LDA band structures, exhibiting total magnetic moment
$M$=1.95 $\mu_B$ that is only accidentally near an integer value, 
are shown in Fig. \ref{fmband}. 
The five Co $3d$ bands and three  $2p$ bands of each O constitute
an entanglement of 17 hybridized bands for each spin, 
with total $p-d$ bandwidths of 10 eV and 11 eV for the majority and minority
bands respectively.
The strong hybridization throughout the bands is evident in the 
corresponding total DOS and accompanying 
atom-projected DOS, shown in Fig. \ref{dos}.
The crystal field splitting of 2 eV, identified from the density of 
states, between $t_{2g}$ and $e_g$ manifolds
is only a little higher than the exchange splitting 1.3 eV.
As noted earlier, these $t_{2g}$ and $e_g$ designations are broken down
by the $4/m$ symmetry of the Co site.
The electronic structure shows clear quasi-two-dimensionality,
consistent with resistivity measurement by Matsuno {\it et al.}

In Fig. \ref{fmband}, the $pd\sigma$ and $pd\sigma^*$ bands
(for each of $d_{x^2-y^2}$ and $d_{3z^2-r^2}$ orbitals) are highlighted.
The more dispersive bands (darker, or red)
arise from bonding and antibonding interactions
between Co $d_{x^2-y^2}$ and PO $p_\sigma$ states.
The bonding and antibonding bands have a separation
of 10 eV at the $X$ point, this total width arising from a combination of
the differences between the $t_{2g}$ and $2p$ site energies, and the
hybridization between them. 
The dispersion of the antibonding (upper) band can be described roughly 
by effective hopping amplitude $t$ 
= 0.40 eV (majority) and $t$ = 0.53 eV (minority).
The behavior of the $d_{x^2-y^2}$ band is similar to that observed 
in the high $T_c$ superconductor
La$_2$CuO$_4$, in which this antibonding band
plays a central role,\cite{mattheiss,pickett} but it is mostly unoccupied here.
However, with the lower number of $3d$ electrons the cobaltate electronic structure is 
influenced more strongly by the $t_{2g}$ manifold, 
which forms relatively flat bands, the top of which hovers around the Fermi energy.
In the minority bands a van Hove singularity lies at $E_F$ 
at the $Z$ point, perhaps contributing to the positioning of the Fermi level and
hence to the total moment.  

The less dispersive highlighted bands (green, or lighter), lying in the 
0 - 2 eV range, 
arise from the antibonding interaction of Co $d_{3z^2-r^2}$ and AO $p_\sigma$ states.
The bonding band of this pair consists of a remarkably flat band at -4 eV for 
each spin direction.
%but the bottom band also has some $d_{3z^2-r^2}$ character.

\subsection{Magnetic Tendency}
\begin{table}[bt]
\caption{Co $3d$ orbital (Mullikan) occupancy in LDA, where $M$=1.95 $\mu_B$. 
 The difference of occupancies between 
 both spin channels is directly related with contribution of each orbital
 to spin magnetic moment,
 which can be seen to be spread over all five $3d$ orbitals.}
\begin{center}
\begin{tabular}{lcccccc}\hline\hline
          &\multicolumn{3}{c}{$t_{2g}$}&~&\multicolumn{2}{c}{$e_g$} \\
                                                   \cline{2-4}\cline{6-7}  
          &\multicolumn{2}{c}{$E_{g}$} & $B_{2g}$ &~&
          $A_g$ & $B_{1g}$ \\\cline{2-3}
          & $xz$  & $yz$  & $xy$  &~& $3z^2-r^2$~ &~ $x^2-y^2$    \\\hline
 majority & ~~1.00~~  & ~~1.00~~  & ~~1.00~~ &~&0.56       & 0.67         \\
 minority    & 0.68  & 0.68 & 0.54 &~& 0.36     & 0.32         \\\hline
 difference~~& 0.32  & 0.32 & 0.46 &~& 0.20     & 0.35    \\ \hline\hline
 \end{tabular}
\end{center}
\label{table1}
\end{table}

The FM state is favored energetically over the nonmagnetic state by 0.37 eV/Co,
similar to the value given by Matsuno {\it et al.} 
The spin magnetic moments are 1.95 total, 1.52 from Co and 0.22 from each PO
(in units of $\mu_B$).
The large magnetic moment for PO, due to strong hybridization 
with Co $3d$ bands, has been observed previously in Li$_2$CuO$_2$\cite{weht} and a few
other cuprates.
The Mullikan decomposition of O charges as well as the band filling of related
bands indicates that AO is consistent with its formal designation 
O$^{2-}$, whereas PO contributes
considerably to the conduction bands and cannot be considered fully ionic.
%provided by the FPLO method, shows PO(-0.67), and AO(-0.87),
%suggesting that PO has more covalent character
%whereas AO has more ionic character.
%Contrast with LS ($t_{2g}^5$; S=1/2), which has been
%observed in formal Co$^{4+}$ system, the Co magnetic moment 
%may be understood by IS ($t_{2g}^4 e_{g}^1$; S=3/2).
According to the Co $3d$ orbital occupancies 
given in Table \ref{table1}, however, every $3d$ orbital contributes 
to the magnetic moment due to itineracy.  The variation from average
contribution occurs in two states: 
$d_{xy}$ has 40\% larger, and $d_{3z^2 -r^2}$ orbital 40\% smaller,
contributions than the average contribution to the moment.
%The total occupation of the $d$ state shows 6.81, much larger value than
%5 as expected from formal Co$^{4+}$ configuration.
%Compared with total occupation of about 7 in a formal Co$^{3+}$ system 
%LaCoO$_3$,\cite{korotin} it implies that the system has strong  
%metallic/covalent character.
The strong itinerant character (all $3d$ orbitals are neither fully occupied
nor fully unoccupied)
explains why no Jahn-Teller distortion
is observed in Sr$_2$CoO$_4$.

%the system should be understood intermediate state between
%metallic/covalent and ionic pictures.
%Considering the ionic radii of Co(+0:1.52; +4:0.69) and 
%O(+0:0.48; -2:1.24) (in~\AA),\cite{atom}
%in the case Co ions in the Co-O$_2$ layer are close each other
%enough for the direct Coulomb interaction to exclude 
%the superexchange mechanism, resulting in ferromagnetic behavior.

\begin{figure}[tbp]
\rotatebox{-90}{\resizebox{7cm}{8cm}{\includegraphics{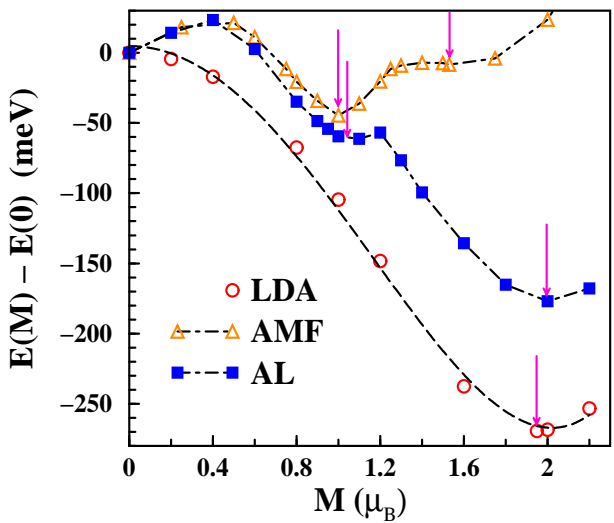}}}
\caption{(Color online) Fixed spin moment calculations in LDA and
 both LDA+U schemes at $U_{c}$=2.5 eV.
 The arrows pinpoint (meta)stable states.
 The dashed line for LDA FSM indicates fitting line with
 $E(M)-E(0)= \varepsilon_0 - \alpha M^2 + \beta M^4$,
 where $\varepsilon_0$=5 meV, $\alpha$=133 meV/$\mu_B^2$, 
 and $\beta$=16 meV/$\mu_B^4$.}
\label{fsm}
\end{figure}

%It is worthwhile to note sharp peaks of the $t_{2g}$ minority, 
%due to flat band, at $\pm$0.3 eV, indicating strong Stoner instability.
%The ferromagnetic behavior stems from the flat band.
%The flat band mechanism for ferromagnetism has been discussed for 
%MgCNi$_3$ \cite{rosner} and calcium pnictides \cite{cap}.
%(Although there are much sharper peaks of $d_{3z^2-r^2}$ near -4 eV,
%it is too away from E$_F$ to affect ferromagnetic instability.)

The fixed spin moment method\cite{fsm} is applied in a following subsection
to probe the magnetic behavior.  In the LDA fixed spin moment calculations
the low moment region is given
by $E(M)-E(0) \approx  - \alpha M^2 + \beta M^4$
with constants  $\alpha$=133 meV/$\mu_B^2$,
$\beta$=16 meV/$\mu_B^4$.
The Stoner-enhanced susceptibility is given by 
\begin{eqnarray}
   \chi=\chi_{0}/[1-N(0)I]
\end{eqnarray}
where the bare susceptibility $\chi_0=2\mu_{B}^{2}N(0)$ and
$N(0)$ is the single-spin density of states at the Fermi level.
Also in the low-$M$ limit one obtains formally $\alpha=(1/2)\chi^{-1}$, giving
the Stoner interaction $I$ = 1.17($\pm$0.05) eV.
With this value $IN(0)$$\approx$ 1.7, giving a very strong Stoner instability 
of the nonmagnetic phase that is numerically similar to that of nonmagnetic Fe.

\subsection{Fermiology}
\begin{figure}[tbp]
\flushleft
{\resizebox{3.85cm}{2.55cm}{\includegraphics{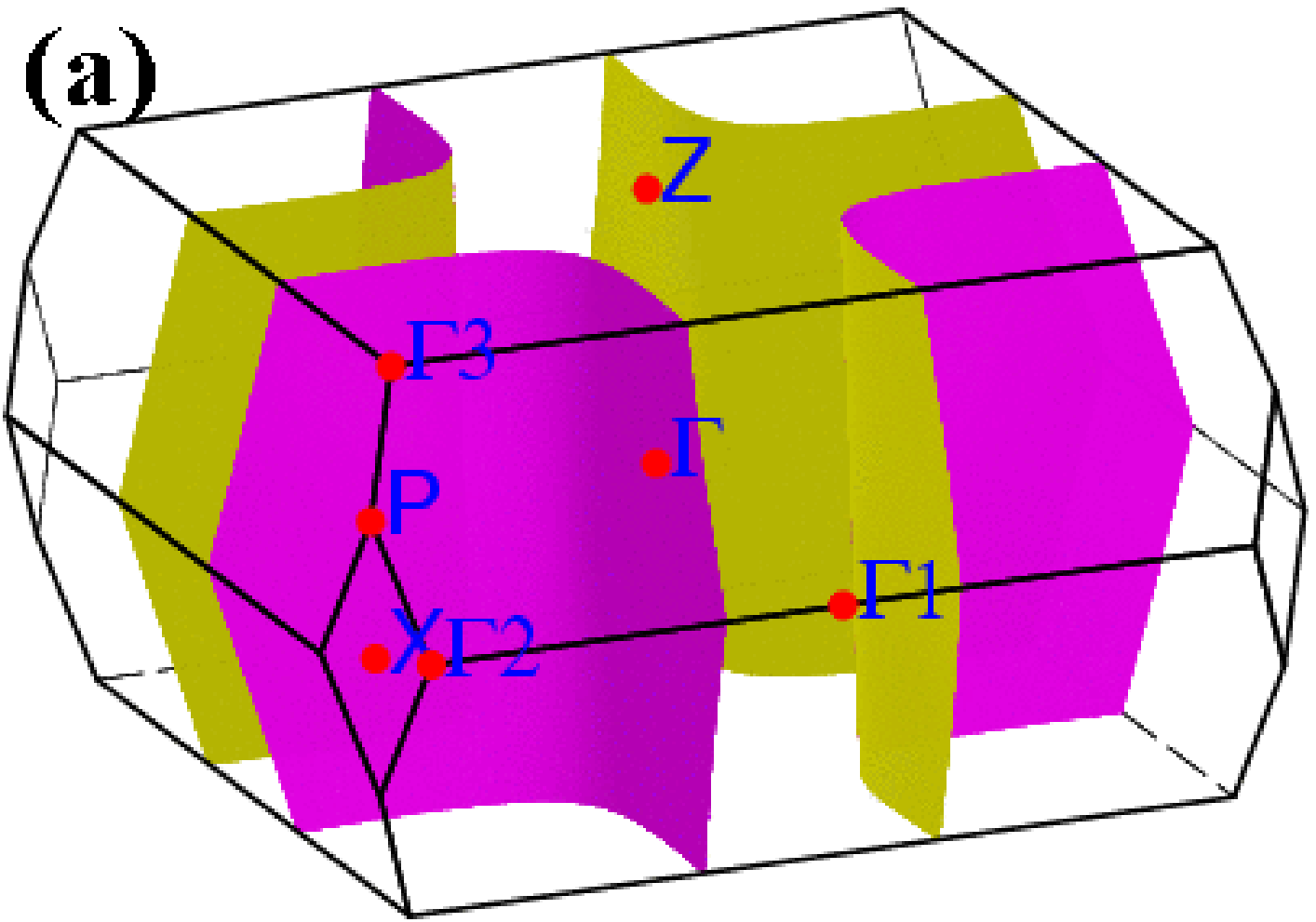}}}
\flushright 
\vskip -29mm
{\resizebox{3.85cm}{2.55cm}{\includegraphics{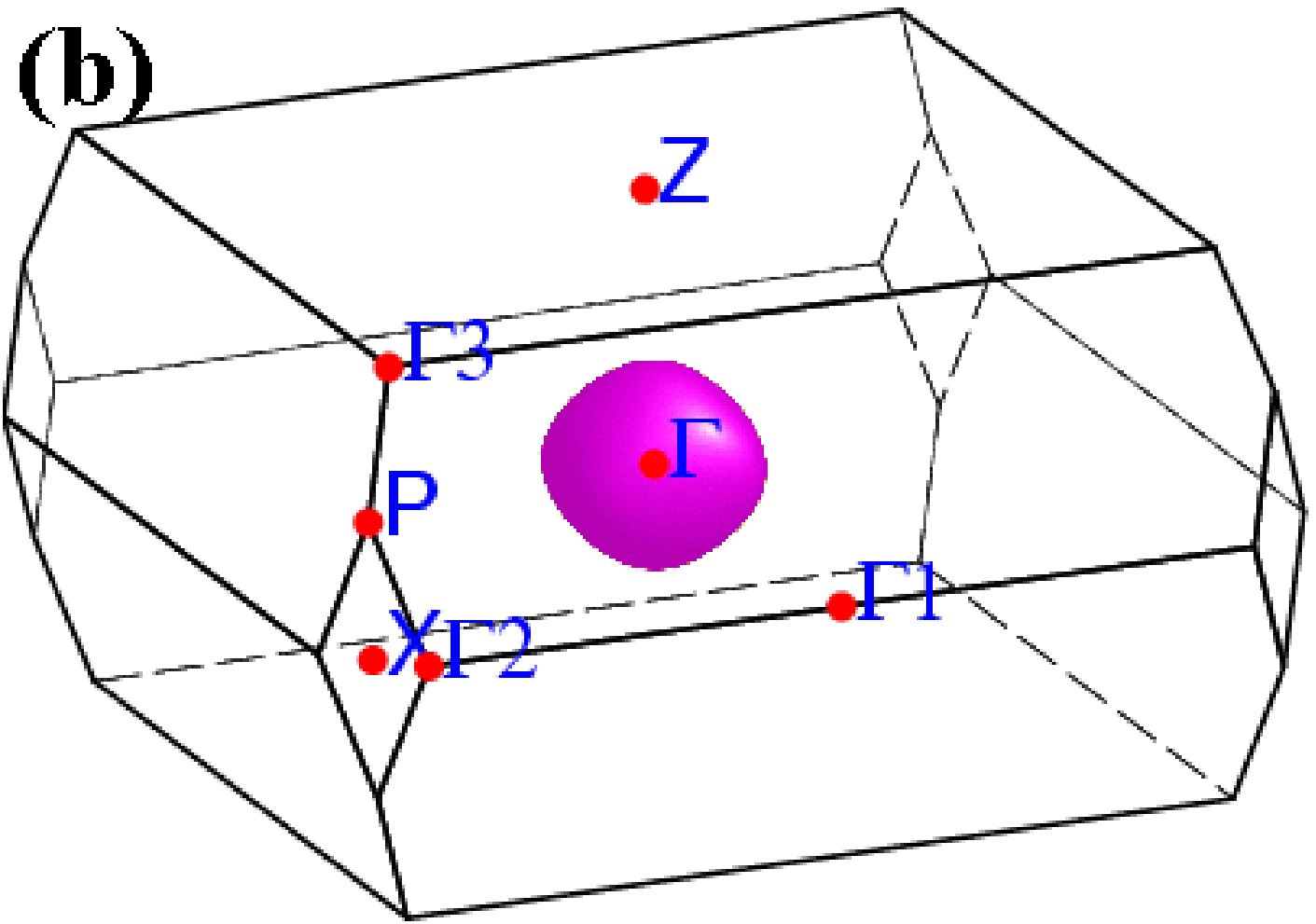}}}
\vskip 4mm
\flushleft
{\resizebox{3.85cm}{2.55cm}{\includegraphics{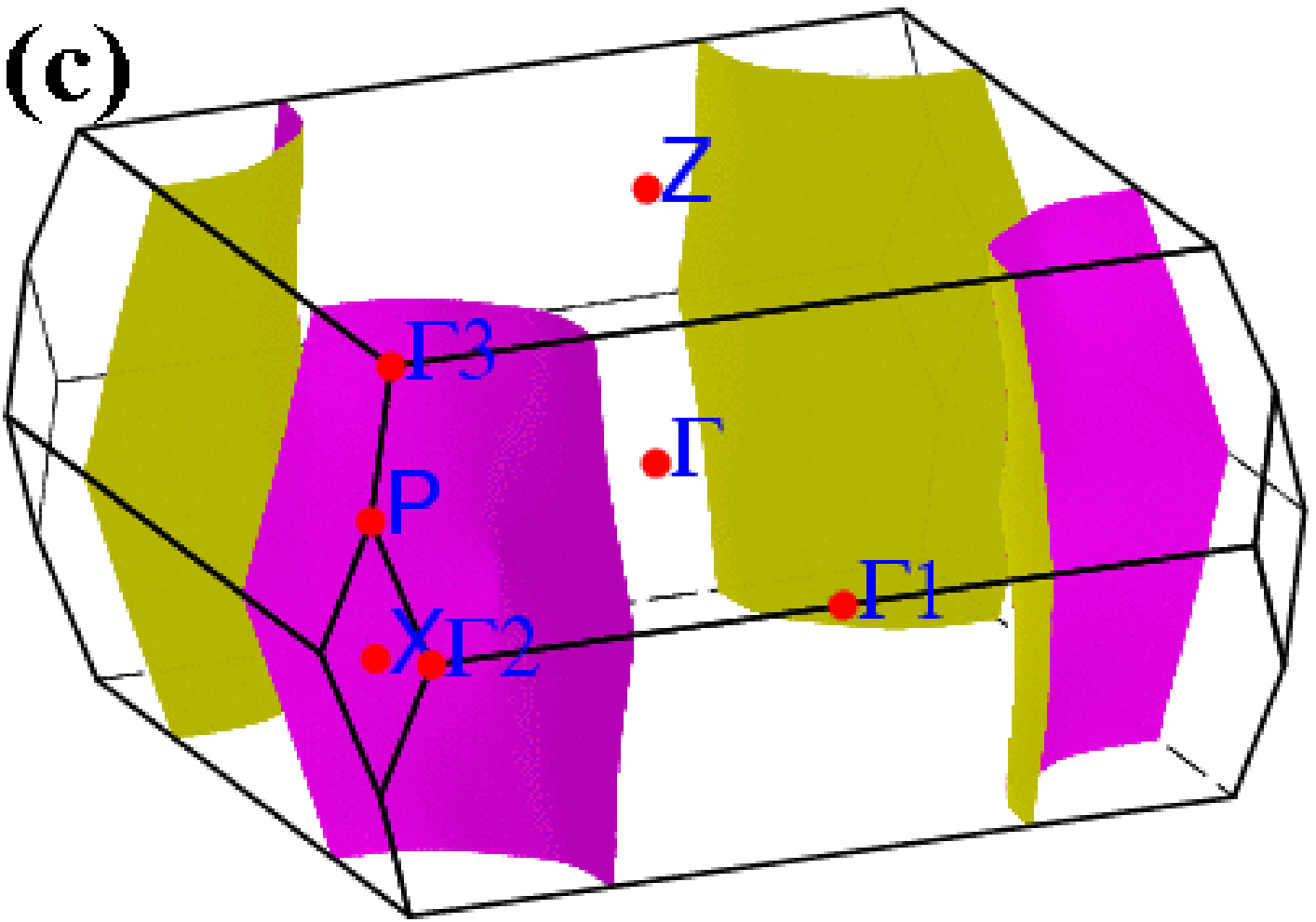}}}
\flushright
\vskip -29mm
{\resizebox{3.85cm}{2.55cm}{\includegraphics{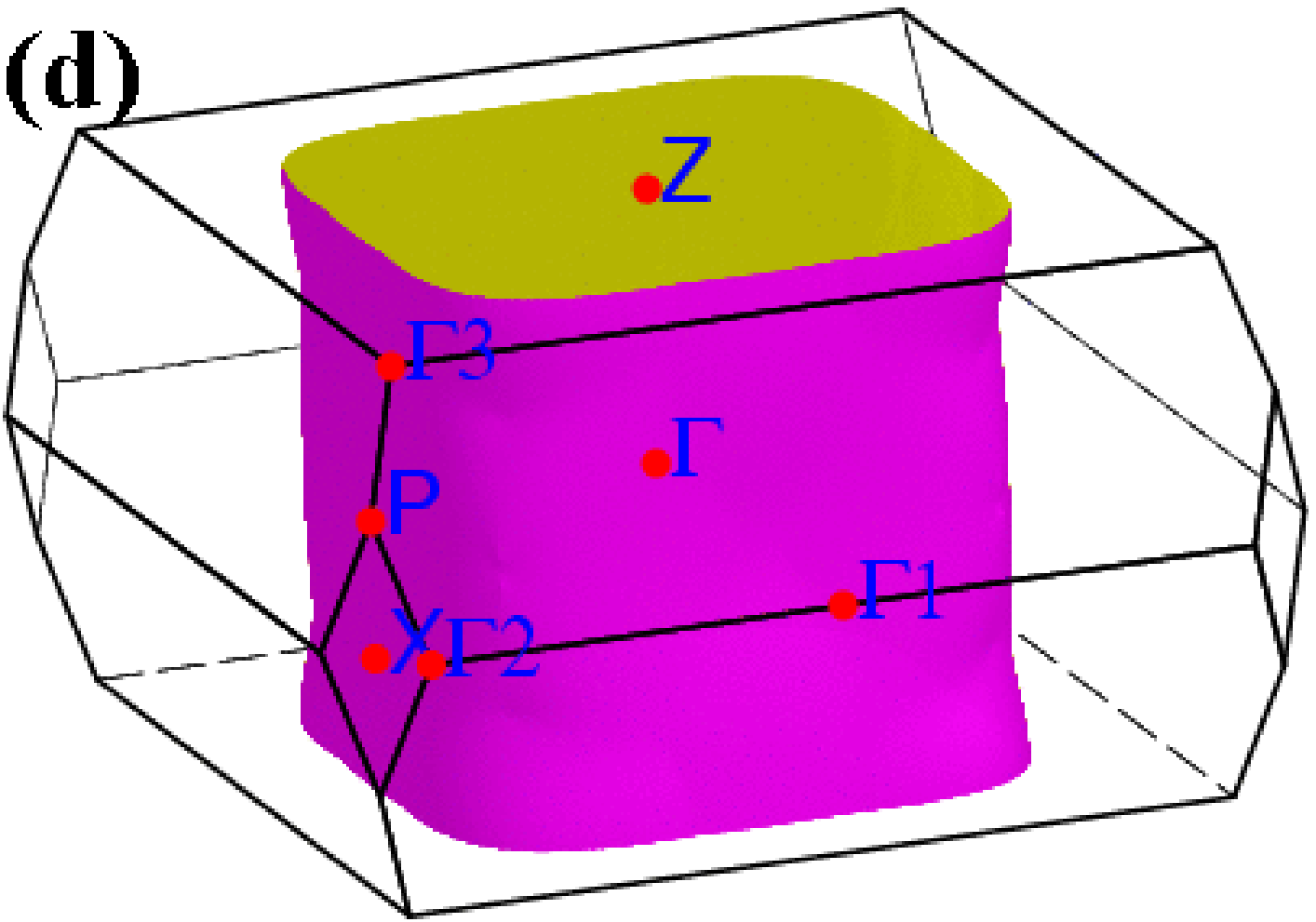}}}
\flushleft
{\resizebox{3.85cm}{2.55cm}{\includegraphics{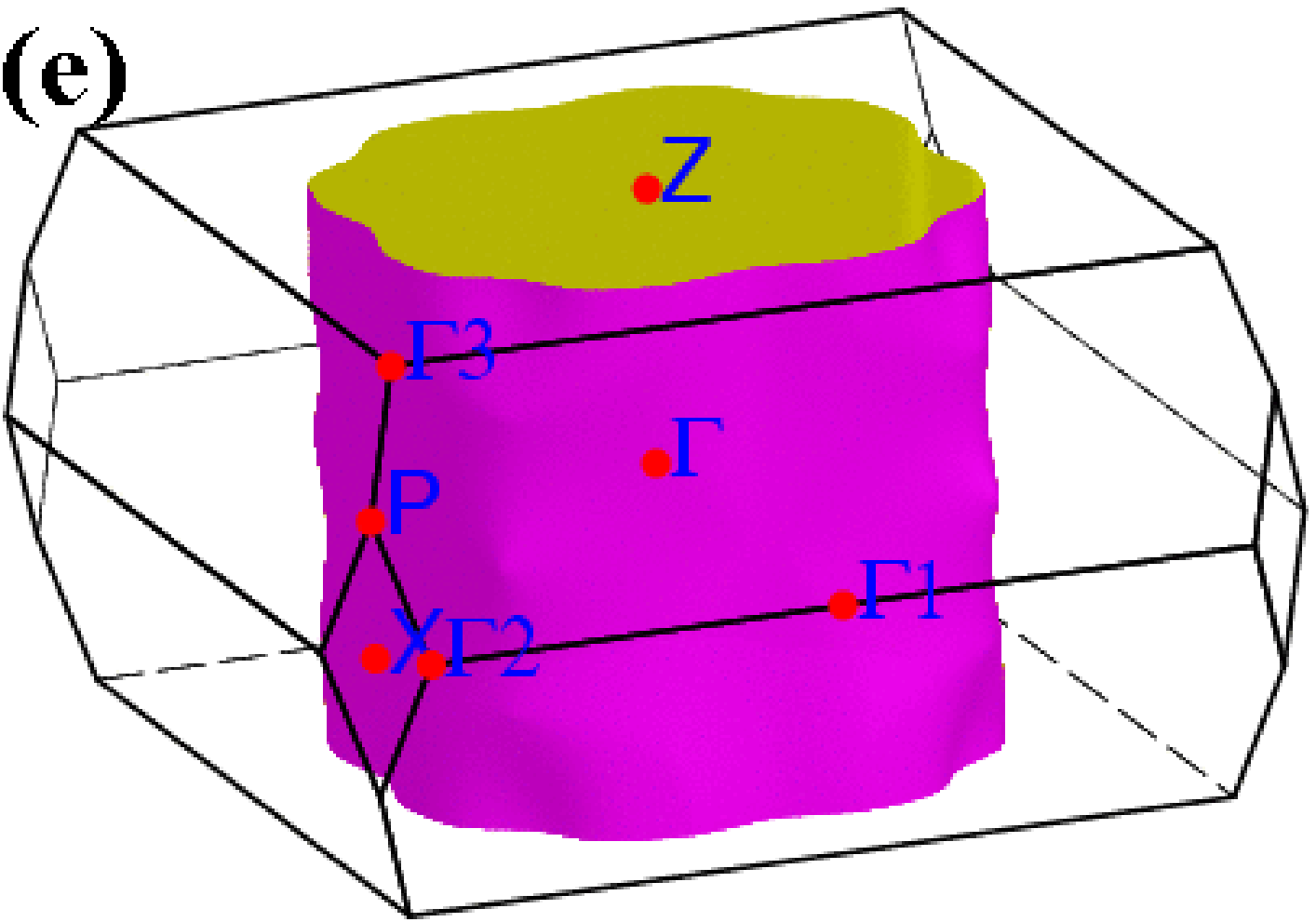}}}
\flushright
\vskip -29mm
{\resizebox{3.85cm}{2.55cm}{\includegraphics{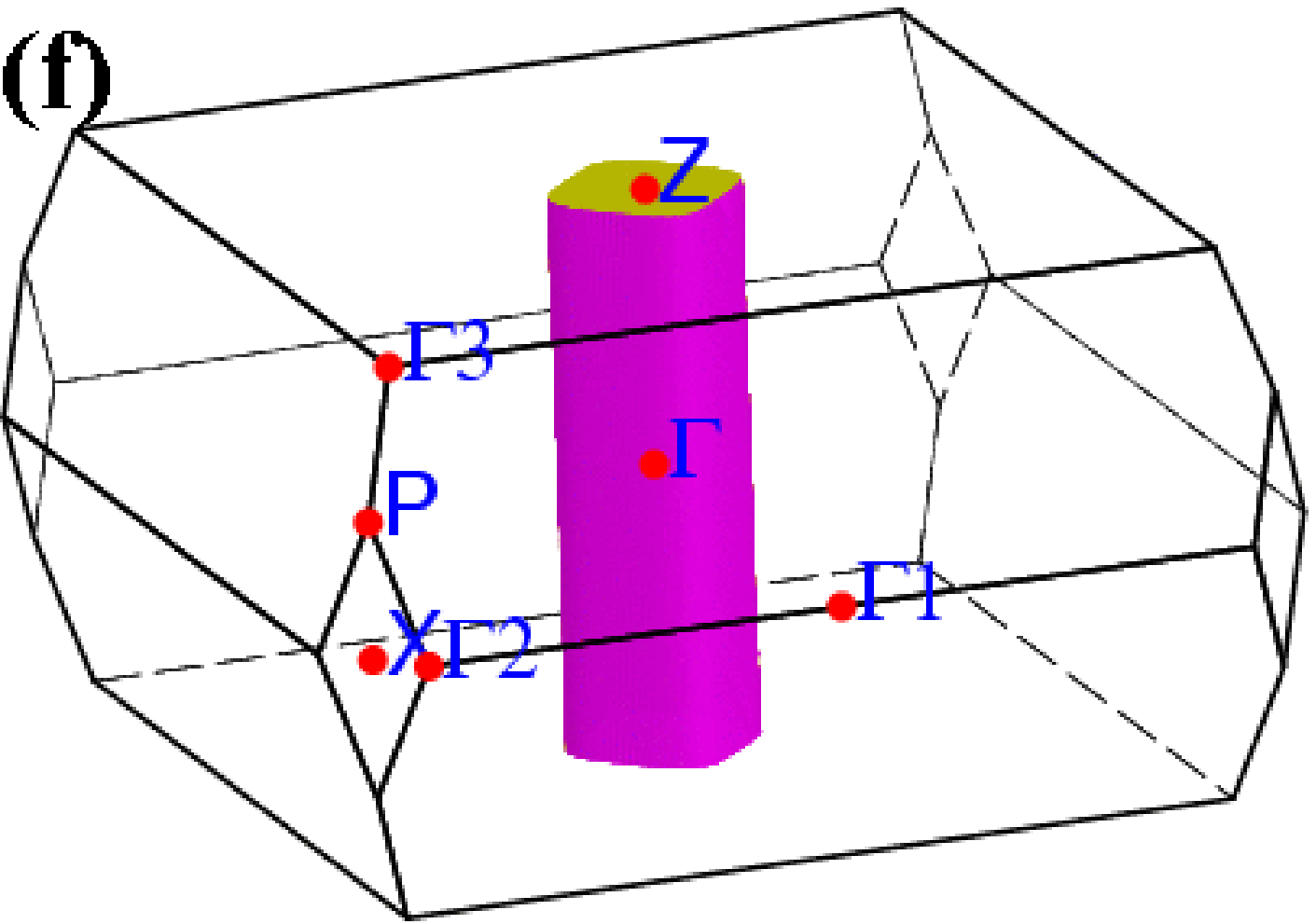}}}
\vskip 4mm
\caption{(Color online) LDA FM Fermi surfaces, for $M$= 1.95 $\mu_B$.
Surfaces (a) and (b) are from the majority states,
 surfaces (c)-(f) from the minority bands.
 While (a) and (c) contain holes, the others enclose electrons.
 Each surface has mainly (a) $d_{x^2-y^2}$ (0.5 holes), 
 (b) $d_{3z^2-r^2}$ (0.01 electrons), (c) $d_{xz}$ (0.2 holes), 
 (d) $d_{xy}$ (0.4 electrons), (e) $d_{yz}$ (0.4 electrons), 
 and (f) $d_{x^2-y^2}$ (0.02 electrons) characters.
 The number in parentheses says carrier number containing each Fermi
 surface per Co.}
\label{FS}
\end{figure}

The LDA FM Fermi surfaces (FS) pictured in Fig. \ref{FS} consist of 
two sheets from the majority states and four minority sheets,
with very simple geometry 
and strong two-dimensionality.
Except for one sheet that has an ellipsoidal shape (majority
$d_{3z^2-r^2}$ character), the FSs have the shape of rectangular
cylinders with rounded corners. The X-centered sheets contain holes, 
whereas the $\Gamma$-centered surfaces contain electrons.
%Although $d_{xz}$ and $d_{yz}$ states are nearly degenerate,
%both states form FS of very different shape since the states display
%different behavior from each other 
%along the $\Gamma$-$\Gamma_1$-$\Gamma_2$ line.
The FS arising from majority $d_{x^2-y^2}$ (Fig. \ref{FS}(a)) 
has a clear nesting feature, as does the minority hole sheet in
Fig. \ref{FS}(d).  The spanning vectors may
lead to spin-density-wave or/and charge-density-wave 
instabilities.  Intra-surface scattering may also show some nesting features.
%The FS contain roughly total 3.9 holes per Co, explaining 4 holes 
%to be required by Luttinger's theorem and formally Co$^{4+}$ valence 
%state, but the formal valence state is not exactly appropriate 
%because of metallicity and hybridization with oxygen.
For the most part Fermi velocities $v_F$ are in the range of a few 10$^7$ cm/sec
as usual in a metal, but order of magnitude lower velocities occur
along the $\Gamma_3$ to $Z$ line at E$_F$.

\subsection{Magnetocrystalline anisotropy}
\begin{figure}[tbp]
%\vskip 10mm
{\resizebox{8cm}{5cm}{\includegraphics{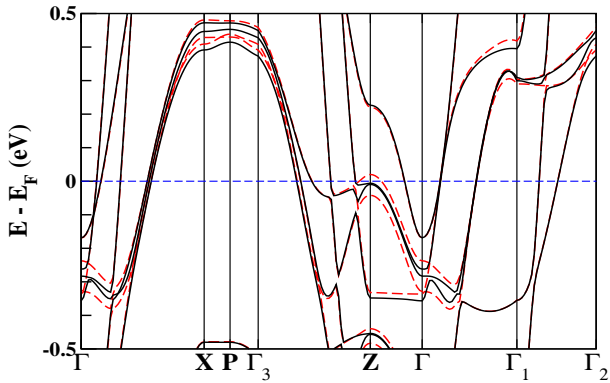}}}
\caption{(Color online) LDA FM band structures from spin-orbit coupling 
 calculation with the quantization axis 
 along $\langle100\rangle$ (solid) and $\langle001\rangle$ 
 (dashed) directions. 
 Note the $\langle001\rangle$ direction is the easy axis.
 The dashed horizontal line denotes the Fermi energy.}
\label{socband}
\end{figure}

We included spin-orbit coupling by using the fully relativistic 
FPLO method\cite{fplo2}
to allow us to study the strong magnetic anisotropy reported\cite{matsuno}
by Matsuno {\it et al.}
The total energies along $\langle100\rangle$ and $\langle001\rangle$
directions were calculated to determine the magnetocrystalline anisotropy
energy (MAE). The energy difference 
$E_{\langle100\rangle} - E_{\langle001\rangle}$=0.57 meV/Co is
consistent with experimentally observed $\langle001\rangle$ easy $c$-axis.
The relativistic calculations give an orbital magnetic moment of
0.08 $\mu_B$ for Co, and negligible values for oxygen ions.
The spin and orbital magnetic moments depend on the spin direction, 
differing by about 5\%.

The experimental MAE can be estimated from field dependent magnetization
along the two directions.\cite{matsuno}
The data show half-saturation at 7 T along 
the $\langle100\rangle$ direction.
Considering the saturated magnetization to be $M_s$=1.8 $\mu_B$/Co,
the MAE from $M_s\cdot B$ is 1.5 meV/Co,
a value three times larger than our calculated result.
Such underestimation in LDA
is common for Co compounds.
The reason is still unsettled, but candidate explanations are the 
poor treatment of
orbital polarization\cite{dresden1} or 
for Hubbard-like correlation\cite{dresden2} in the standard LDA.

Rotating the spin direction induces a change of
band structure, as shown Fig. \ref{socband}.
While the band structure for the moment along $\langle100\rangle$ direction is
essentially that of LDA, there are additional splittings at several points  
for the moment oriented along the easy axis.
The most visible splitting occurs in Co $t_{2g}$ manifold, 
where the spin-orbit splitting spans E$_F$ at the Z point.
Applying SOC with spin directed along the easy axis, the bands split into
primarily $d_{j=5/2, m_j=-3/2}$ character in the upper band, and 
$d_{j=3/2, m_j=1/2}$ in the lower band.
The splittings are 60 meV and 90 meV at the Z and 
$\Gamma$ point respectively.

\section{Inclusion of correlation effect}
\label{ldau}

\begin{figure}[tbp]
\rotatebox{-90}{\resizebox{7cm}{8cm}{\includegraphics{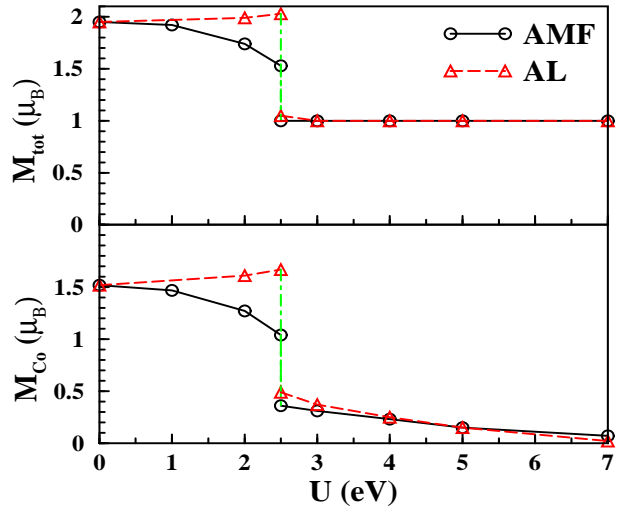}}}
\caption{(Color online)
 Effect of the on-site Coulomb repulsion $U$ on total and Co
 local magnetic moments in both LDA+U schemes.
 At $U_{c}$=2.5 eV, a metal to half metal transition occurs.
 The first-order transition is obtained nearly at the same $U$
 in the both schemes. (In fact, the transition occurs 
 a little higher $U_{c}$ in FLL, but the difference is only less 
 than a few tenth eV.)}
\label{um}
\end{figure}

\begin{figure}[tbp]
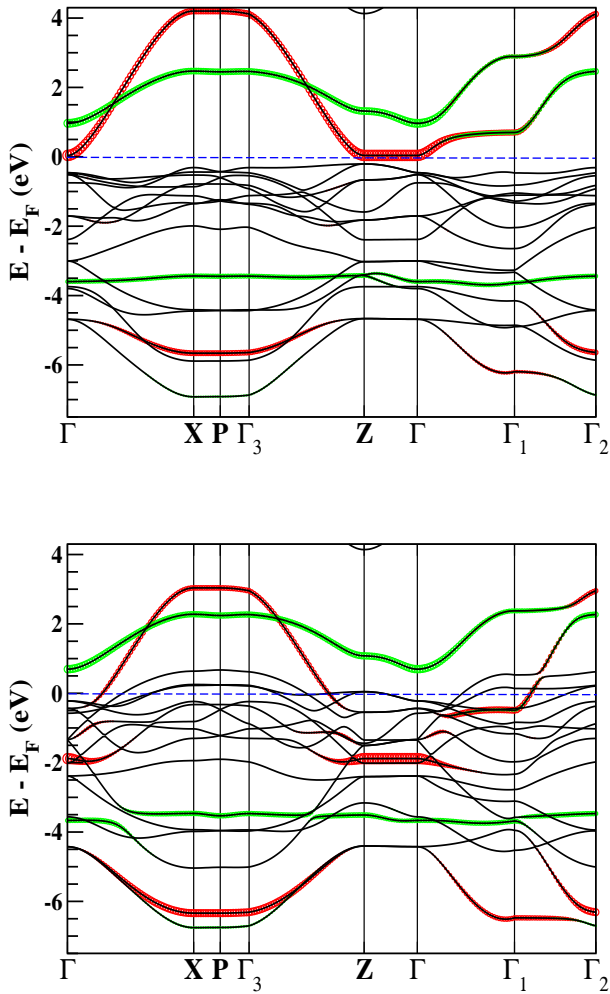

{\resizebox{8cm}{6cm}{\includegraphics{Fig9a.eps}}}
\vskip 11mm
{\resizebox{8cm}{6cm}{\includegraphics{Fig9b.eps}}}
\caption{(Color online) LDA+U FM majority (upper) and
 minority (lower) band structures at $U_{c}$=2.5 eV in the half metallic state
 with $M$=1 $\mu_B$.
 The majority state shows a gap of 0.25 eV.
 The thickened (and colored) lines emphasize Co $d_{3z^2-r^2}$
 (green or light) and $d_{x^2-y^2}$ (red or black) characters.
 The dashed horizontal line denotes the Fermi energy. 
}
\label{fmuband}
\end{figure}

\begin{table}[bt]
\caption{Co $3d$ orbital occupancy in half metallic state, 
 having $M$=1 $\mu_B$, at $U$=2.5 eV. 
 Compared with LDA result shown in Table \ref{table1}, there are two
 remarkable changes in the $e_g$ state; nearly vanishing contribution
 of d$_{3z^2-r^2}$ and negative contribution of d$_{x^2-y^2}$
 (for details, see text).
 It seems to be close to LS state, but the d$_{x^2-y^2}$ minority
 has considerable occupancy due to strong hybridization (itinerary)
 which makes impossible to be called strictly as LS state.
 The total occupation is 6.79.
}
\begin{center}
\begin{tabular}{lcccccc}\hline\hline
          &\multicolumn{3}{c}{$t_{2g}$}&~&\multicolumn{2}{c}{$e_g$} \\
                                                   \cline{2-4}\cline{6-7}
          &\multicolumn{2}{c}{$E_{g}$} & $B_{2g}$ &~&
          $A_g$ & $B_{1g}$ \\\cline{2-3}
          & $xz$  & $yz$  & $xy$  &~& $3z^2-r^2$~ &~ $x^2-y^2$    \\\hline
 majority & ~~1.00~~  & ~~1.00~~  & ~~1.01~~ &~&0.35       & 0.25         \\
 minority    & 0.81  & 0.81 & 0.44 &~& 0.44     & 0.68         \\\hline
 difference~~& 0.19  & 0.19 & 0.57 &~& -0.09     & -0.43   \\ \hline\hline
 \end{tabular}
\end{center}
\label{table2}
\end{table}

\subsection {Metal to Half-metal Transition} 
Since the appropriate value of $U$ in this and other cobaltate systems is unclear, 
we have studied the ground state as $U$ is increased. The moment initially 
increases slightly in FLL, while it decreases slowly to
1.5 $\mu_B$ at $U$=2.5 eV in AMF, as shown in Fig. \ref{um}.
From their separate states just below and at the critical value
$U_{c}$=2.5 eV, beyond this critical 
value the moment drops sharply to 1 $\mu_B$
in both schemes, and then both schemes produce very similar results
in the entire region above $U_{c}$.  Note particularly that this transition starts
from distinct states, but occurs at the same value $U_c$ to the same final state.
Both high-moment and low-moment states can be stabilized in the
calculations at $U_{c}$.  Unlike studies in the Na$_x$CoO$_2$ system,
no discernible hysteretic region could be found at this first-order transition.
This magnetic collapse accompanies a metal to half metal transition,
presumably because there is a particular stability of this half metallic
(HM) FM state since both 
LDA+U schemes transition to it.
The HM state, with a magnetic moment 1 $\mu_B$,
has been observed also by Wang {\it et al.}
As $U$ increases, the Co local magnetic moment decreases to vanishingly
small value by $U\sim$6-7 eV.  The state remains a HM FM, the moment has been
pushed onto the PO ions.  Since the moment on Co has vanished, it is not
surprising that the two LDA+U schemes give the same result.

The microscopic mechanism behind the magnetic collapse induced by the on-site
Coulomb repulsion $U$ can be unraveled from a study of the charge decompositions
in Tables \ref{table1} and \ref{table2}, and comparison of band structures.  
We can compare and contrast the two
viewpoints.  First we   point out that, while the Mullikan charges given 
in these tables are somewhat basis set dependent so their specific magnitude
should not be given undue significance, differences -- whether between orbitals
or between spin-projections -- are more physical.  Then, while these orbital
occupations provide one characterization, we have also provided in Figs.
\ref{fmband} and \ref{fmuband} by the fatbands technique, the bands that one identifies
with $d_{x^2-y^2}$ and $d_{3z^2-r^2}$ character.  Using both viewpoints provides
a more robust interpretation of behavior than either separately.

From the tables one can determine that increasing $U$ does not change the total
$3d$ occupation, but introduces rearrangements in the $3d$ charge and moment.  
From the bands, the main
difference (between FM LDA and HM FM LDA+U) is that the dispersive majority 
$d_{x^2-y^2}$ band has become fully {\it un}occupied, leaving the small gap that results
in half metallicity.
Returning to the charges and moments, what stands out is that the moment
on the $d_{x^2-y^2}$ orbital has flipped, 
from +0.35$\mu_B$ to -0.43$\mu_B$.  The $d_{3z^2-r^2}$
moment has undergone a smaller change in the same manner: 0.20$\mu_B$ to -0.09$\mu_B$.
Thus the net moment on the Co ion, which is $\sim0.5\mu_B$, is the result of 
$\sim +1\mu_B$ in the $t_{2g}$ orbitals (primarily $d_{xy}$) and $\sim-0.5\mu_B$
in the $e_g$ orbitals.
This type of cancellation, in more striking form, has been seen previously in LDA+U results
for LaNiO$_2$,\cite{lanio} which has an unusually {\it low} formal oxidation state for
a nickelate.

\subsection{Fixed Spin Moment Calculation at $U_{c}$}
The results of the previous section show that two magnetic states
coexist at $U_{c}$ in LDA+U, whether one uses the AMF or FLL functional.
Such a coexistence has been already observed
in LDA+U calculations for the sodium cobaltates, where the change at $U_{c}$
corresponds to a charge disproportionation transition.\cite{nacoo}
Here the change is simply in the state of the (single) Co ion; there is no
experimental indication of disproportionation here.
Here we analyze the FSM results at $U_{c}$ for LDA, AMF, and FLL.

The energy versus total magnetic moment behavior,
displayed in Fig. \ref{fsm}, shows very interesting differences as well as
similarities.  Perhaps most interesting is that positions of local
minima occur at (or near) integer values M=0, 1, and 2 (in units of $\mu_B$), 
as if there might be underlying
states with $S_z$ = 0, $\frac{1}{2}$, or 1 pervading the behavior.  The
results of previous sections however established that there is strong $d-p$ 
hybridization which renders integral moments no more favored than other values.
The LDA curve shows simple behavior: a Stoner instability at M=0 and a single 
minimum near (but not precisely at) M=2.  
The LDA+U results show more interesting variation. 

For both AMF and FLL the paramagnetic state is metastable, with energy rising
up to a moment $M\approx 0.5$, and then decreasing very similarly to the minimum at
$M$=1.  Beyond that  point, AMF levels off to a broad flat region $M$=1.3--1.8
beyond which it increases with $M$.  FLL however switches over to
a separate phase, with its minimum at $M$=2.  The E($M$) curve for FLL
seems to consist of two separate parabolas (different phases) with minima
very near $M$=1 and $M$=2.  The M=1 result is integral because a half metallic
phase is encountered (see Fig. \ref{fmuband}) and not due to an $S=1/2$ 
configuration of the Co ion.  Both majority $d_{x^2-y^2}$ and 
$d_{3z^2-r^2}$ become completely unoccupied, leaving a small gap to the
$t_{2g}$ bands.

We must point out that for both AMF and FLL schemes,  the minimum at 
$M$=0 $\mu_B$ does not correspond to a nonmagnetic state.  There is a
moment on Co with magnitude 0.14 $\mu_B$ that
is canceled by magnetization on the PO ion.  The driving force for favoring this
low, canceling moment phase over a nonmagnetic state is not clear.  Possibly
an antiferromagnetic result, requiring a doubled unit cell, would be a lower
energy M=0 solution.

\begin{figure}[tbp]
\rotatebox{-90}{\resizebox{7cm}{8cm}{\includegraphics{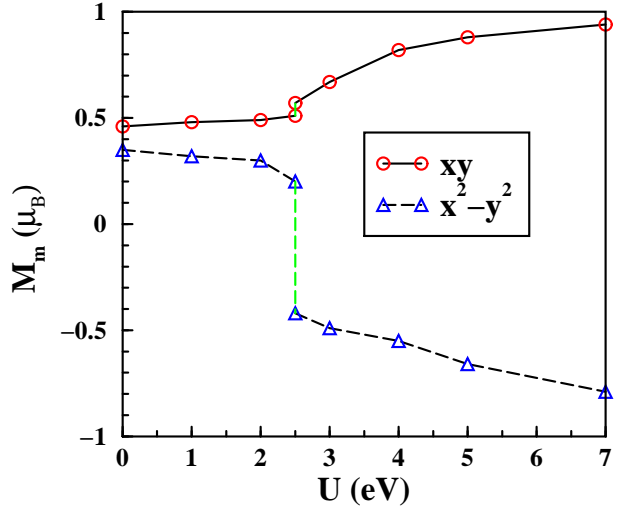}}}
\caption{(Color online) Effect of strength of $U$ on the orbital-projected Co 
magnetic moment $M_{m}$, 
 defined by difference between majority and minority occupancies,
 of Co $d_{x^2-y^2}$ and $d_{xy}$ states in AMF scheme. 
 (Results for the FLL scheme are similar.)
 In the large $U$ limit an on-site ``singlet" type cancellation of  
 moments occurs in this $m=\pm 2$ channel (see Fig. \ref{um} and text).}
\label{occ}
\end{figure}

\subsection {Strong Interaction Regime}
In Sr$_2$CoO$_4$ correlation effects should not be very strong it seems, since LDA
already gives a FM state that seems consistent with one of the experimental
reports.
However, in a strong correlated regime, but not beyond realistic range of $U$,
there is another interesting feature.
Figure \ref{occ} shows the $U$-dependent orbital-projected
local magnetic moment $M_m$, defined by the difference between majority
and minority occupancies, of Co $d_{x^2-y^2}$ and $d_{xy}$ states.
Upon increasing $U$, the $d_{xy}$ minority state loses electrons, while
the $d_{x^2-y^2}$ minority state gains electrons.
Beyond $U=U_c$ = 2.5 eV, $d_{xy}$ and $d_{x^2-y^2}$ have large positive and negative
local magnetic moments respectively, 
leading to an on-site ``singlet" type cancellation within this $|m|=2$ channel.
This type of cancellation, but within the $e_g$ manifold, has been seen 
already in our previous results
for LaNiO$_2$.\cite{lanio}
In contrast with LaNiO$_2$, which shows the cancellation 
in the total magnetic moment and Ni$^{1+}$$\rightarrow$Ni$^{2+}$ conversion,
the moment of the system remains unchanged because of large magnetic
moments on planar oxygen ions.

\section{Summary}
Synthesis of the high formal oxidation state compound Sr$_2$CoO$_4$, bulk materials by
high-pressure high-temperature techniques
and films by pulsed laser deposition (PLD),
have led to a high Curie temperature metallic ferromagnet that introduces new
transition metal oxide physics and may be useful in spin electronics devices.
We have provided an in-depth study of the electronic and magnetic structure of
this compound, looking specifically into the combined effects of correlation
on the $3d$ orbitals and strong hybridization with O $2p$ states.  

Within LDA Sr$_2$CoO$_4$ is metallic with a ferromagnetic moment near $2\mu_B$,
close to the saturation magnetization reported for the PLD films.
Application of the two commonly used LDA+U functionals reveals several
surprises.  As $U$ is increased from zero, the two functionals produce changes
in the moment of opposite sign up to the critical value $U_c$=2.5 eV.  Beyond
%U_c$ there is a first-order transition for both functionals to the {\it same}
half metallic ferromagnetic phase with moment 1$\mu_B$.  Within this phase,
increasing the value of $U$ has the effect of pushing the moment completely
off the Co and onto the planar O ions by $U\sim$ 6-7 eV, while the total 
moment remains fixed at 1$\mu_B$.

With $U$ fixed at $U_c$=2.5 eV, fixed spin moment calculations were carried 
out for both LDA+U functionals, and compared with the corresponding LDA result.
Both LDA+U schemes behaved similarly out to a minimum at 1$\mu_B$ (the half
metallic state).  Beyond this the two functionals departed in their behavior,
with the FLL scheme jumping to a new state with minimum very near 2$\mu_B$,
not half metallic and much the same as the LDA minimum (1.95$\mu_B$).
In the stronger-interacting regime $U \geq 3$ eV (which may not be appropriate
for Sr$_2$CoO$_4$), the correlation described by the LDA+U approach leads to
oppositely directed moments on the $d_{x^2-y^2}$ and $d_{xy}$ orbitals,
reflecting the strong difference in hybridization of these two orbitals.

\section{Acknowledgments}
We acknowledge illuminating conversations with M. Richter for 
magnetic anisotropy energy and K. Koepernik for technical assistance.
This work was supported by DOE grant DE-FG03-01ER45876 and DOE's
Computational Materials Science Network.  W.E.P. acknowledges the
stimulating influence of DOE's Stockpile Stewardship Academic Alliance Program.

\end{document}